\documentclass[aps,prl,twocolumn,reprint,longbibliography]{revtex4-1}
\usepackage{epsfig}
\usepackage{color}
\usepackage{amsmath}
\usepackage{amsfonts}
\usepackage{natbib}
\usepackage{graphicx}
\usepackage{subfigure}
\usepackage{multirow} 
\usepackage{lipsum}
\usepackage{afterpage}
\usepackage{placeins}
\usepackage{float}
\usepackage{braket}
\usepackage[normalem]{ulem}

\newcommand{\rb}{\mathbf{r}}

\newcommand{\avg}[1]{\left<#1\right>}
\newcommand{\len}[1]{\left|#1\right|}

\newcommand{\para}[1]{\left(#1\right)}

\newcommand{\Mb}{\ensuremath{\mathcal{M}}}

\begin{document}

\title{Octahedral tilting and $B$-site off-centering in halide perovskites are not coupled}

\author{Colin M. Hylton-Farrington}
\author{Richard C. Remsing}
\email[]{rick.remsing@rutgers.edu}
\affiliation{Department of Chemistry and Chemical Biology, Rutgers University, Piscataway, NJ 08854, USA}

\begin{abstract}

Metal halide perovskites show exceptional potential for solar energy, thermoelectrics, catalysis, and other photochemical technologies, with performance rooted in electronic structure-driven properties. 
In $ABX_3$ halide perovskites, localized and often aspherical local electron densities from $B$-site lone pairs or polarizable $X^-$ anions can distort the lattice.
However, the links among electronic structure fluctuations and distortions like tilting of the $BX_6$ octahedra and off-centering of the $B$-site from the center of its octahedron are not fully understood.
Using group theory and ab initio molecular dynamics, we quantify how lone pairs, halide polarization, off-centering, and octahedral tilting interact in the cubic phase Cs$B$Br$_3$, with $B$ = Pb, Sn, and Ge.
We find that lone pair-induced off-centering and octahedral tilting are symmetry-decoupled.
Instead, stereochemical lone pair expression of the $B$-site ion is correlated to octahedral tilting through the 
propensity of the $B$-site to form a transient, partial covalent bond with the surrounding halide ions that stiffens octahedral tilting modes. 
These results link local electronic asymmetry to structural fluctuations and suggest that dynamic modulation of electronic symmetry offers a pathway to control functional properties in halide perovskites.

\end{abstract}

\maketitle

\raggedbottom

Lattice dynamics and their coupling to electronic structure control many important processes in metal halide perovskites. 
Halide perovskites exhibit soft, anharmonic phonon modes that have attracted significant interest because of their critical role in governing properties such as low thermal conductivity, dynamic disorder, carrier scattering, exciton binding and fine structure, ionic transport, and phase stability~\cite{ubaid2024, acharyya2020, eames2015, mayers2018, swift2023, becker2018bright, biffi2023excitons, weinberg2023size, park2024theoretical, maity2024}.
In cubic halide perovskites of the form $ABX_3$, two lattice distortions that have garnered much attention are octahedral tilting~\cite{yang2017spontaneous},  
which occurs when alternating $BX_6$ octahedra rotate around an axis, and off-centering of the $B$-site from the center of its octahedron~\cite{fabini2016}. 
Octahedral tilting increases the electronic gap, such that dynamic tilting increases fluctuations in the gap and alters absorption properties like the Urbach energy~\cite{skettrup1978urbach,greeff1995anomalous,yang2017spontaneous,hyltonfarrington2024}.
Local structural changes from octahedral tilting modify orbital overlap and symmetry relevant to carrier effective masses and transport~\cite{schilcher2021significance,mayers2018,schilcher2023correlated,vonhoff2025analysis}. 
If the $B$-site has a lone pair, the lone pair will localize to one side of the ion (due to $s-p$ mixing, for example), and interactions with the surrounding halide ions produce an instability in which the $B$-site is displaced away from the center of the metal-halide octahedron~\cite{waghmare2003first,bovzin2010entropically,fabini2016,Dutta:2021aa}. 
This off-centering, or emphanisis, significantly affects the optical properties of perovskites, such as photoluminescence~\cite{fabini2016}, and can suppress their thermal conductivity~\cite{xie2020}. 
Despite the importance of octahedral tilting and off-centering modes in the photophysics and thermoelectric behavior of halide perovskites, their physical origin is not fully understood, even at the level of whether or not these two modes are linked.
Previous work has postulated that off-centering is coupled to octahedral tilting~\cite{gao2021,smith2015}.
However, a $B$-site lone pair and the resulting emphanisis are not a prerequisite for halide perovskites to exhibit octahedral tilting~\cite{tkachenko2021empty,egger2024}.
Moreover, DFT calculations suggest that off-centering and octahedral tilting might even compete~\cite{radha2018distortion}. 
As a result, the question remains: how is $B$-site lone pair expression linked to octahedral tilting, if at all? 
In this communication, we answer this question for cubic Cs$B$Br$_3$ perovskites using a combination of molecular dynamics simulations and group theory-based analysis of coupled electronic and nuclear fluctuations.
We tune the amount of lone pair expression by varying the $B$-site in cubic Cs$B$Br$_3$ from Pb to Sn to Ge and quantify how lone pair expression couples to off-centering and octahedral tilting. 
By quantifying the local electronic symmetry of the $B$-site, we show that off-centering and octahedral tilting exhibit incommensurate symmetries, such that \emph{lone pair expression is not coupled to octahedral tilting}. 
Using the insights uncovered here by our group theory-based analysis, we suggest strategies for tuning these uncoupled low-energy instabilities in cubic halide perovskites through the local electronic structure of the component ions. 
%

\section{Results}

 \begin{figure*}[tb]
   {\includegraphics[width=0.99\textwidth]{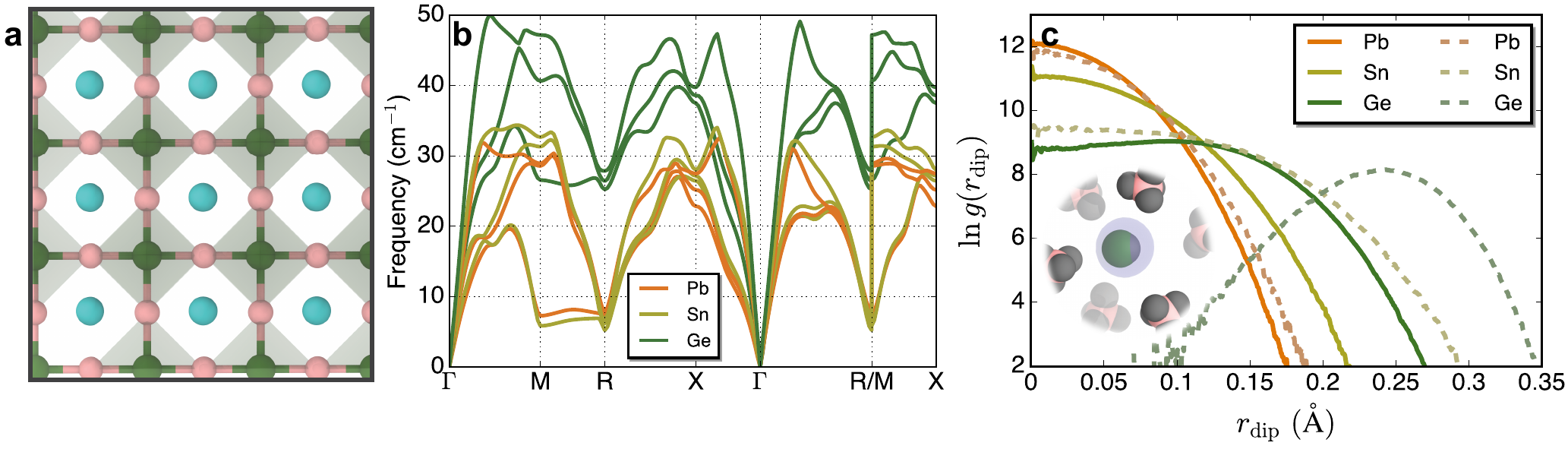}}   
  \caption{\textbf{Structure and lattice dynamics of cubic Cs$B$Br$_3$.}
  \textbf{a} Snapshot of the cubic CsGeBr$_3$ supercell of length $4L$ used in simulations. Structures for CsSnBr$_3$ and CsPbBr$_3$ are similar but have different lengths, $L$. Cs is cyan, Br is pink, and Ge is green. 
  \textbf{b} Acoustic branches of the phonon dispersion relations.
  \textbf{c} Natural logarithm of the radial distribution function for the distance between $B$-site cation and its lone pair MLWFC, $r_{\rm dip}$, which quantifies the expression of the $B$-site lone pair. 
  Dashed lines correspond to $g(r_{\rm dip})$ for systems modeled at the same unit cell length and temperature. 
 The inset shows a simulation snapshot of a Ge ion (green), its MLWFC (gray), and the corresponding MLWF (blue isosurface). Also shown are surrounding Br ions (pink) and their MLWFCs (gray). }
  \label{fig:1}
\end{figure*}

%
At high temperatures, $ABX_3$ perovskites typically adopt a cubic crystal structure, Fig.~\ref{fig:1}a.
However, this cubic structure is only obtained upon averaging; the instantaneous structure fluctuates away from the average cubic structure.
The unstable nature of the cubic structure is evidenced by imaginary frequencies in the 0~K phonon dispersion curves~\cite{yang2017spontaneous,zhou2020, huang2014, kashikar2024}.
The presence of imaginary frequencies indicates mechanical instability, which, when allowed to relax, causes the perovskite to fluctuate between locally distorted states. 
At finite temperature, the phonon dispersion curves are renormalized by their anharmonicity, resulting in nearly flat, dispersionless bands between the M and R points in the Brillouin zone, Fig.~\ref{fig:1}b, which are attributed to octahedral tilting. 
This behavior is observed for CsPbBr$_3$, CsSnBr$_3$, and CsGeBr$_3$, albeit with different frequencies.
The frequencies of the three acoustic branches shown in Fig.~\ref{fig:1}b for all three systems are of similar frequencies for CsPbBr$_3$ and CsSnBr$_3$, while the frequencies of CsGeBr$_3$ are generally higher, reflecting a stiffer crystal.
The differences among the three perovskites can in part be attributed to variations in the electronic structure of the $B$-site cations due to increased $s-p$ mixing, in addition to their different sizes. 
To characterize the local electronic structure and its fluctuations, we use maximally localized Wannier function centers (MLWFCs) as a point representation of the electron density, which we have found to be a reasonable approximation for localized charge densities~\cite{hyltonfarrington2024,dhattarwal2024}.
We quantify the amount of lone pair expression through the ion-MLWFC pair distribution function, $g(r_{\rm dip})$,
where $r_{\rm dip}$ is the distance between the center of the ion and the MLWFC (the length of the $B$-MLWFC dipole). 
It is clear that the width of the peak in $g(r_{\rm dip})$ increases from Pb to Sn to Ge.  
Additionally, the maximum in $g(r_{\rm dip})$ shifts to a non-zero value in the case of Ge, indicating substantial lone pair expression. 
The variation in lone pair expression among the three perovskites considered here enables us to systematically identify correlations (or lack thereof) among lone pair expression, $B$-site off-centering, and octahedral tilting.
The amount of lone pair expression is not the only difference between the three systems studied here; 
they differ in unit cell length ($L$) and temperature.
To isolate the effects of $B$-site lone pair expression, we model CsPbBr$_3$, CsSnBr$_3$, and CsGeBr$_3$ all at the same unit cell length, 
corresponding to that of cubic CsPbBr$_3$, as well as at the same temperature of 550~K. 
The amount of lone pair expression in these model systems follows expectations, $\rm{Pb}<\rm{Sn}<\rm{Ge}$, as evidenced by $g(r_{\rm dip})$, (Fig.~\ref{fig:1}c, dashed lines).
In fact, the larger unit cell significantly increases the amount of lone pair expression for Sn and Ge. 
The qualitative trends discussed below are also exhibited by the systems at the experimentally determined unit cell size, as discussed further in the SI and shown in Fig.~S4-S7,
and we focus on the model systems with the same $L$ throughout the rest of the main text unless otherwise noted.
%

 \begin{figure}[tb]
   {\includegraphics[width=0.49\textwidth]{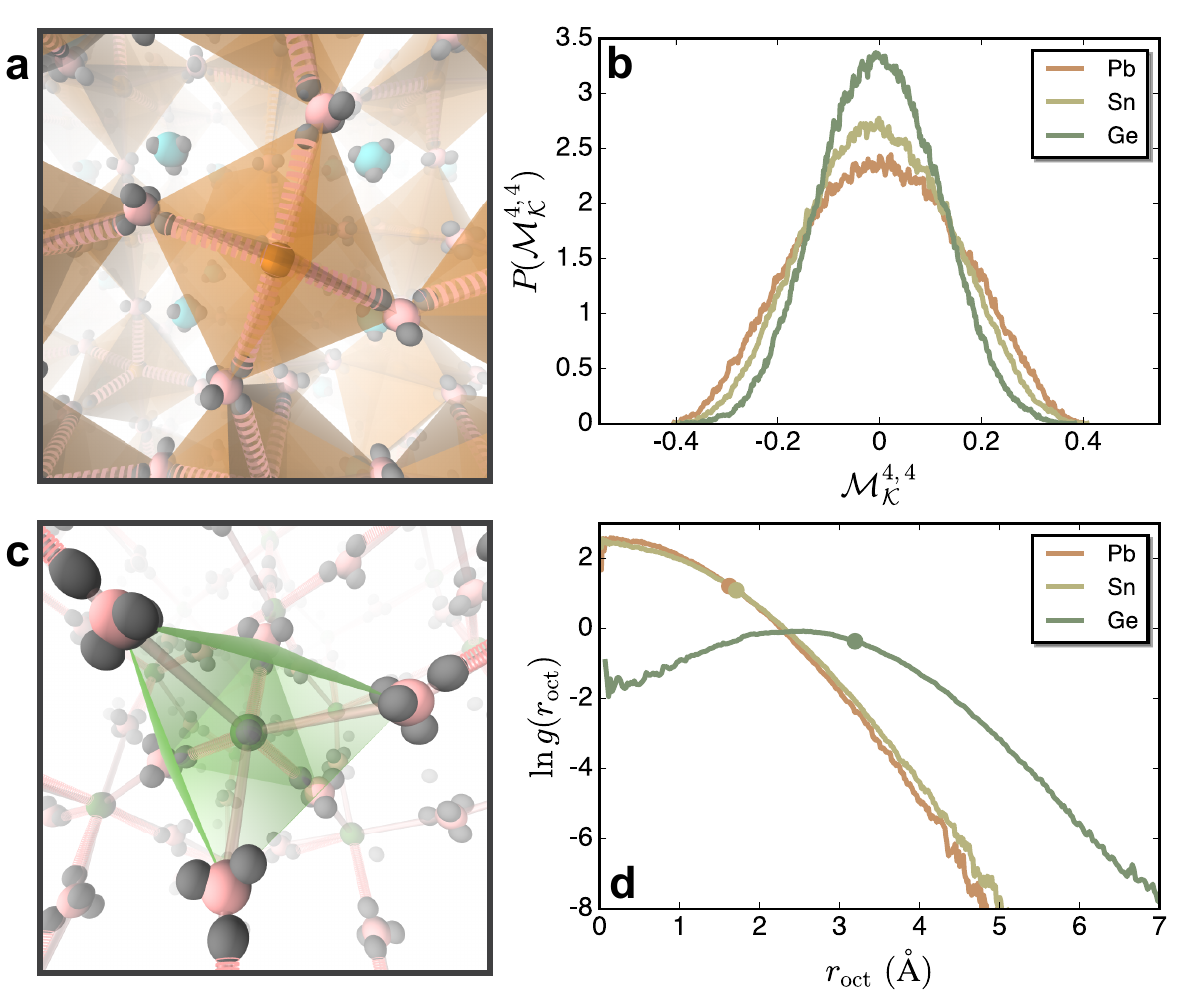}}   
  \caption{\textbf{Octahedral tilting and $B$-site off-centering are of different symmetries.}
  \textbf{a} Snapshot of CsPbBr$_3$ undergoing octahedral tilting. Cs is cyan, Pb and octahedral units are colored orange, and Br is pink. 
  In-plane bromides are connected to Sn with dashed cylinders, and MLWFCs are shown as gray spheres.  
  \textbf{b} Probability distribution of the rotor function quantifying octahedral tilting, $P(\Mb_{\mathcal{K}}^{4,4})$, where deviations away from zero result from octahedral tilting.
 \textbf{c} Snapshot of CsGeBr$_3$ in a configuration exhibiting $B$-site off-centering. 
One set of symmetry equivalent bromides is connected to Ge with dashed cylinders, while the other set is connected to Ge with transparent solid cylinders.
 MLWFCs are also shown to further illustrate symmetry-equivalent ions and their local electron density, as well as the orientation of the Ge-lone pair dipole.
 \textbf{d} Radial distribution function, $g(r_{\rm oct})$, between the $B$-site and the center of the $B$Br$_6$ octahedron, $r_{\rm oct}$. Data points indicate the average distance, $\avg{r_{\rm oct}}$.
    } 
  \label{fig:2}
\end{figure}

%
Our quantification of octahedral tilting and off-centering is rooted in group theory,
mainly using rotor functions as order parameters to quantify atomic and electronic symmetry fluctuations~\cite{lyndenbell1994,klein1983,hyltonfarrington2024}.
A rotor function, 
\begin{equation}
\mathcal{M}_{\xi}^{\ell,m} = \frac{1}{N}\sum_{i=1}^{N}\mathcal{S}_{\xi}^{\ell,m}\para{\Theta_{i}},
\label{eq:rotor}
\end{equation}
is a function of the orientation of a collection of vectors, usually around a central atom, whose value indicates how well the collective orientation overlaps with the appropriate symmetry-adapted function, ${\mathcal S}_{\xi}^{\ell,m}\para{\Theta_{i}}$,
which is defined by angular momentum ($\ell$) and index $m$, bounded by $0 \le m \le 2l$.
The symmetry-adapted function belongs to the irreducible representation (irrep) of the site symmetry of the central atom of interest and can be used to understand group-to-subgroup transformations straightforwardly.
We use these functions to quantify fluctuations in local nuclear and electronic structures accompanying octahedral tilting and $B$-site off-centering.
In the cubic structure, the six bromides of the $B$Br$_6$ octahedron are equivalent, and distortions away from this structure reduce the number of symmetry equivalent bromides.
Octahedral tilting separates the bromides into four in-plane and two axial bromides, where only the in-plane bromides participate in the rotation that results in tilting, Fig.~\ref{fig:2}a.
Tilting couples to changes in the collective orientation of the bromide electron density, with the four equivalent bromides orienting an MLWFC toward the $B$-site.
These collective fluctuations away from the average cubic structure result from the incommensurate symmetries of the site and the local electron density~\cite{hyltonfarrington2024}.
%

 \begin{figure*}[tb]
   {\includegraphics[width=0.9\textwidth]{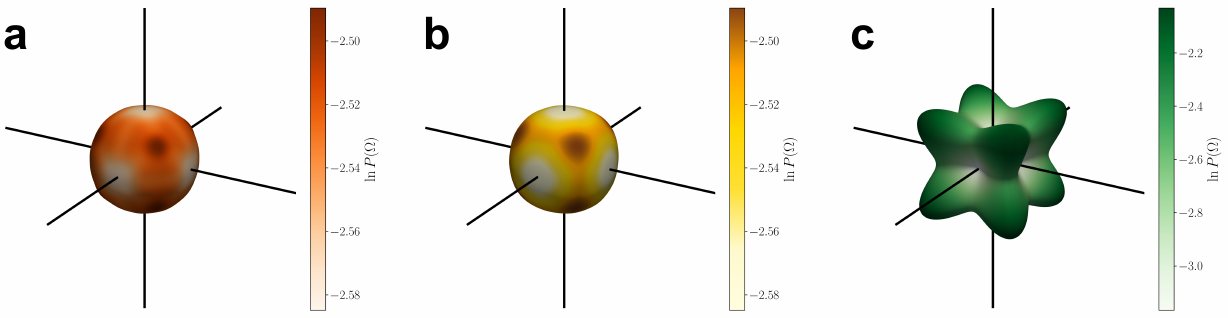}}   
  \caption{\textbf{$B$-site lone pairs point along off-diagonal [111] axes.}
  Orientational probability distribution functions for the $B$-site ion-MLWFC unit vector for \textbf{a} CsPbBr$_3$, \textbf{b} CsSnBr$_3$, and \textbf{c} CsGeBr$_3$.
  } 
  \label{fig:3}
\end{figure*}

%
Octahedral tilting transforms the environment of the $B$-site cation from $O_h$ to $C_{4h}$, such that changes in symmetry can be monitored with an order parameter that belongs to the $T_{1g}$ irrep.
Here, we use the rotor function ${\mathcal M}_{\mathcal K}^{4,4}$ with $B$-Br vectors as inputs; ${\mathcal M}_{\mathcal K}^{4,4}=0$ in the cubic structure and $\len{{\mathcal M}_{\mathcal K}^{4,4}}>0$ if there is octahedral tilting.
The probability distribution of this rotor function, $P(\Mb^{4,4}_{\mathcal{K}})$, 
narrows when changing the $B$-site from Pb to Sn to Ge,
which indicates decreasing amounts of octahedral tilting from CsPbBr$_3$ to CsSnBr$_3$ to CsGeBr$_3$, Fig.~\ref{fig:2}b.
Therefore, increasing lone pair expression from Pb to Ge appears to decrease the propensity of halide perovskites to exhibit octahedral tilting, yet the precise nature of this coupling is not known. 
Stereochemical lone pair expression has been linked to off-centering of the $B$-site. 
The predominant off-centering mode is along the $[111]$ direction, which lowers the symmetry of the $B$-site from $O_h$ to $C_{3v}$ as it moves toward one of the faces of the Br$_6$ octahedron~\cite{fabini2016}.
This reduction in site-symmetry creates two sets of three equivalent bromides around the $B$-site, Fig.~\ref{fig:2}c.
To quantify the amount of off-centering in each system, we computed the radial distribution function (RDF) between the position of the $B$-site and the center of the octahedron, $g(r_{\rm oct})$, where
\begin{equation}
r_{\rm oct} = \len{\rb_B - \sum_{i=1}^6 \rb_{X_i}},
\end{equation}
$\rb_B$ is the position of the $B$-site cation, and $\rb_{X_i}$ is the position of the $i$th halide in the octahedron.
This off-centering RDF, $g(r_{\rm oct})$, indicates that Pb and Sn off-center similarly, though $\avg{r_{\rm oct}}$ is slightly larger with Sn, while Ge off-centers significantly further, Fig.~\ref{fig:2}d.
Therefore, increasing lone pair expression, from Pb to Ge, increases the system's propensity to undergo $B$-site off-centering.
%

 \begin{figure*}[tb]
   {\includegraphics[width=.99\textwidth]{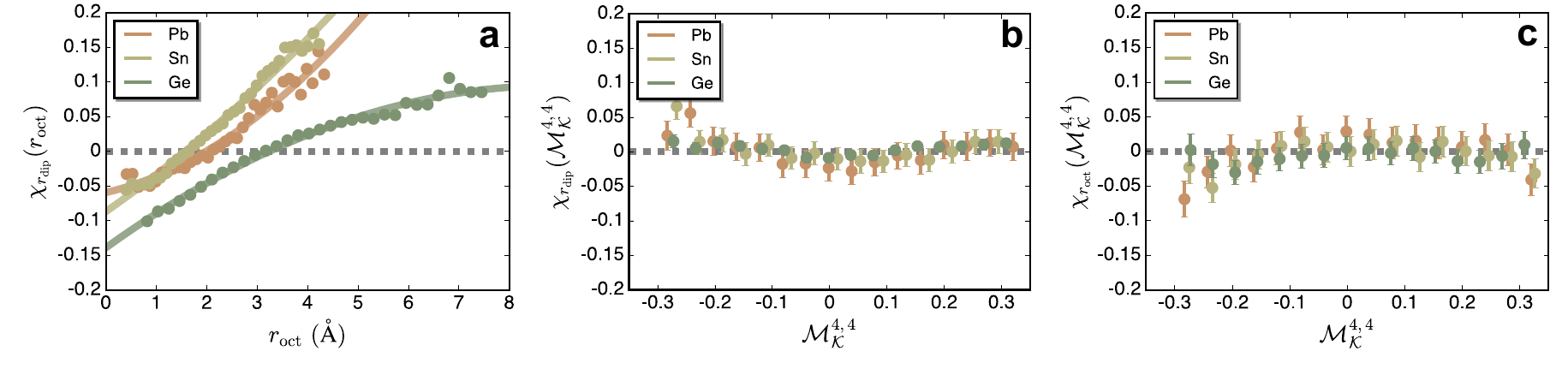}}   
  \caption{\textbf{Stereochemical lone pair expression is coupled to off-centering but not octahedral tilting.}
  \textbf{a} Average fractional distance between the $B$-site cation and its MLWFC ($r_{\rm dip}$) conditioned on the distance between the $B$-site and the center of the octahedron ($r_{\rm oct}$)
  \textbf{b} Average fractional $B$-site dipole conditioned on the octahedral tilting order parameter.
  \textbf{c} Average fractional off-centering conditioned on the octahedral tilting order parameter.
  Error bars indicate one standard deviation. 
  }
  \label{fig:4}
\end{figure*}

%
Off-centering of the $B$-site can occur through distortions with two different symmetries.
The first distortion, discussed above, lowers the $B$-site symmetry from $O_h$ to $C_{3v}$ and corresponds to a polarization along a $[111]$ (off-diagonal) axis.
This off-diagonal mode separates the six Br ions into two groups of three equivalent Br ions that displace in opposing off-centering directions.
The second distortion lowers the $B$-site symmetry from $O_h$ to $C_{4v}$ and corresponds to polarization along a $[100]$ (Cartesian) axis. 
The change in symmetry accompanying this Cartesian distortion separates the six Br ions of the octahedron into a set of four equivalent in-plane Br ions and two axial Br ions that displace in opposing directions. 
Based on symmetry, the Cartesian, $[100]$ distortion is allowed to couple to octahedral tilting while the off-diagonal, $[111]$ distortion cannot. 
To quantify which polarization mode accompanies off-centering, we computed the probability as a function of the $B$-MLWFC vector orientation,  $P(\Omega)$, where $\Omega = (\phi,\theta)$ is the solid angle.
To highlight the directional preferences, we show $\ln P(\Omega)$ in Fig.~\ref{fig:3}, where the value 
of $\ln P(\Omega)$ is indicated by color, and the distortion of the surface away from a sphere is proportional to $P(\Omega)$.
The lone pair of the $B$-site should point toward one of the faces of the $B$Br$_6$ octahedron to minimize electrostatic repulsion with the bromides, and therefore we anticipate that off-centering induces polarization along the off-diagonal $[111]$ axis.
The orientational probability distributions, $P(\Omega)$, indicate that all three systems orient their $B$-site lone pair along the $[111]$ axis. 
All probabilities exhibit minima (white) along the $[100]$ directions and maxima (darker colors) along the $[111]$ directions, such that orienting the $B$-site lone pair along the $[111]$ direction is preferred while orienting the lone pair along $[100]$ is unfavorable.
This preferential orientation along the $[111]$ direction is in agreement with experimental work that attributed the asymmetry of X-ray pair distribution functions to the transient presence of rhombohedral ($R3m$) distortions in the average cubic perovskite structure~\cite{fabini2016}.
The relatively large lone pair expression of Ge manifests in significantly larger differences in the probability distribution, as compared to the Pb and Sn systems, such that differences between the minima and maxima in $P(\Omega)$ are an order of magnitude larger in the Ge system.
Overall, larger lone pair expression favors more off-centering and an increasing preference to point the lone pair toward the face of the Br$_6$ octahedron, consistent with expectations due to minimizing electrostatic repulsions. 
%

 \begin{figure*}[tb]
   {
   \includegraphics[width=0.9\textwidth]{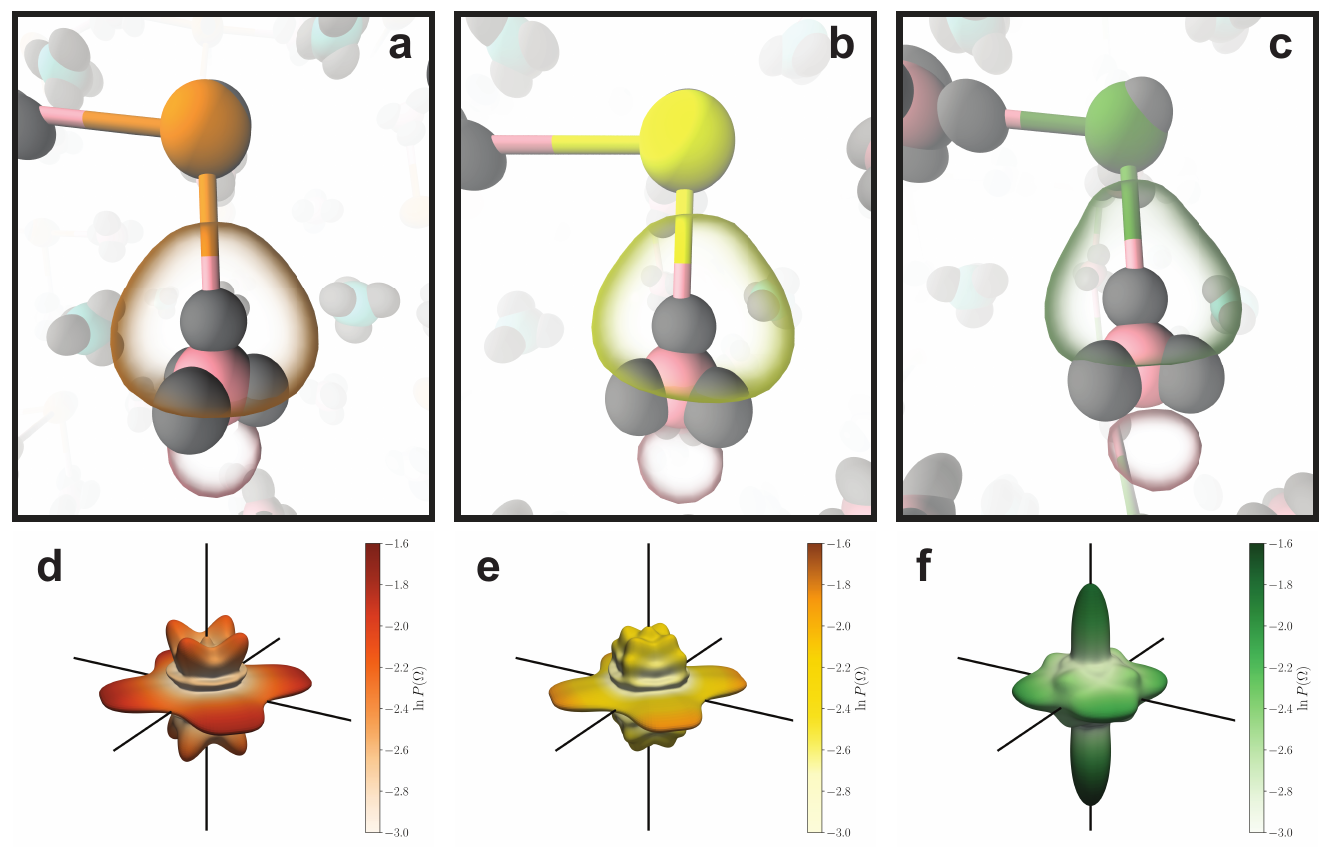}
   }   
  \caption{\textbf{Bonding between bromide and the $B$-site.}
  \textbf{a-c} Snapshots showing the MLWF connecting bromide to \textbf{a} Pb, \textbf{b} Sn, and \textbf{c} Ge. 
  MLWFs are drawn at isosurface values of 0.108592 and 0.052187~Bohr$^{-3}$ for bonding and antibonding surfaces.  \textbf{d-f} Orientational probability distribution functions for the bromide MLWFCs of \textbf{d} CsPbBr$_3$, \textbf{e} CsSnBr$_3$, and \textbf{f} CsGeBr$_3$.}
  \label{fig:5}
\end{figure*}

%
The incommensurate $B$-site symmetries of octahedral tilting and off-centering along the $[111]$ direction suggest that lone pair expression and octahedral tilting are not correlated. 
To further quantify any coupling among $B$-site off-centering, lone pair expression, and octahedral tilting, 
we computed combinations of conditional probability distributions, shown in Fig.~S1-S3. 
To illustrate the main findings, we plot the fractional deviation of various averages in Fig.~\ref{fig:3},
which are defined as
\begin{equation}
\chi_x(y) \equiv \frac{\left<x(y)\right>}{\left<x\right>}-1,
\end{equation}
where $\avg{x(y)} = \int dx x P(x|y)$ and $\avg{x} = \int dy \avg{x(y)} = \int dy \int dx x P(x|y)$.
We first examine $\chi_{r_{\rm dip}}(r_{\rm oct})$, which quantifies how the average distance between the $B$-site ion and its lone pair MLWFC changes with the displacement of the $B$-site away from its ideal position in the cubic crystal structure. 
We find clear correlations between $r_{\rm dip}$ and $r_{\rm oct}$ for all three materials. 
However, the precise nature of the correlation is different for each, as illustrated by the quadratic fits to each data set (solid lines).
When $B=$~Pb, $r_{\rm dip}$ is roughly quadratic with $r_{\rm oct}$ and exhibits positive curvature. 
When $B=$~Ge, $r_{\rm dip}$ is also quadratic with $r_{\rm oct}$ but exhibits negative curvature. 
In contrast, $r_{\rm dip}$ is approximately linear with $r_{\rm oct}$ when $B=$~Sn. 
We attribute the different trends to the different coordination environments of the $B$-site. 
For Pb, tilting dominates over off-centering, such that the coordination shell of Pb mainly consists of 4 equivalent bromides,
while off-centering dominates tilting for Ge and its coordination shell mainly has two sets of 3 equivalent bromides.
Sn displays similar amounts of tilting and off-centering, such that $\avg{r_{\rm dip}(r_{\rm oct})}$ behaves in a manner intermediate between Pb and Ge.
Conditional averages of $r_{\rm dip}$ and $r_{\rm oct}$ as a function of the octahedral tilting order parameter $\mathcal M^{4,4}_{\mathcal K}$ show little to no correlation, as evidenced by $\chi_{r_{\rm dip}}(\mathcal M^{4,4}_{\mathcal K})$ and  $\chi_{r_{\rm oct}}(\mathcal M^{4,4}_{\mathcal K})$ in Fig.~\ref{fig:3}b,c.
As a result, both stereochemical lone pair expression and off-centering are not correlated to octahedral tilting. 
This lack of correlation is expected from the symmetry-based arguments discussed above; tilting and ($[111]$) off-centering are transformations with incommensurate symmetries.
Because off-centering and lone pair expression are coupled, the same symmetry arguments imply that lone pair expression is also not directly coupled to octahedral tilting. 
If lone pair expression and off-centering are not correlated to octahedral tilting, why does the amount of octahedral tilting decrease from Pb to Sn to Ge? 
The decrease in tilting results from changes in Br--$B$-site bonding. 
The differences in Br--$B$-site bonding can be qualitatively observed in the Br MLWF closest to the $B$-site.
The MLWF in the Pb system exhibits minimal directionality, consistent with the small propensity for the MLWFC to align with the $z$-axis. 
In contrast, the MLWF in the Ge system exhibits a `teardrop' shape consistent with partial covalent bonding between Br and Ge. 
The Sn system is in between these two limits. 
We can quantify the changes in the local Br$^-$ electron densities through the orientational distribution functions of the bromide MLWFCs, Fig.~\ref{fig:5}d-f.
 For each distribution, there are two major contributions:
 (1) off-diagonal peaks with four-fold symmetry that are consistent with the site symmetry when tilting~\cite{hyltonfarrington2024}, and
 (2) peaks along the $\pm z$-axis that correspond to a Br MLWFC pointing directly toward the $B$-site,
 consistent with a partial Br--$B$ bond.
From Pb to Sn to Ge, the strength of the Br--$B$ bond increases, evidenced by the increase in the peaks along the $z$-direction.
 The high propensity for the MLWFC to align with the local $z$-axis and point toward a bromide in CsGeBr$_3$ is a signature of the partial covalent bonding observed qualitatively in the MLWF isosurfaces. 
The observed evolution of MLWF shape and tilt stiffness from Pb to Sn to Ge is consistent with the electronic-structure picture developed by Papoian and Hoffmann: increasing $s-p$ mixing promotes partial covalent (directional) bonding of the anion-cation pair and converts nonbonding lone-pair density into bonding character, thereby stiffening the relevant lattice modes~\cite{A.-Papoian:2000aa,papoian2001electron,ienco2001electron,PAPOIAN19988}.
The MLWF shapes we observe correspond to a continuous transition between localized lone-pair character and partial multicenter bonding; this electronic transition explains the simultaneous increase in $r_{\rm dip}$ and increased Br--$B$ bonding directionality seen for Ge.
This framework rationalizes why enhanced lone-pair expression (larger $r_{\rm dip}$) in Ge coexists with reduced octahedral tilting; the stronger directional Br--$B$ bonding raises the energetic cost of octahedral rotations. 
 Octahedral fluctuations in halide perovskites are dictated not by lone pair expression but by the strength or stiffness of the Br--$B$-site bond.
Octahedral tilting and $B$-site off-centering occur in the presence of this bond, and therefore the relative propensity for either distortion is tied to the strength of the Br--$B$-site bond.
Fluctuations between off-centering configurations require breaking this bond, and the nature and extent of octahedral tilting fluctuations are dependent on the bond strength.
As a result, a dynamical competition should exist between octahedral tilting and $B$-site off-centering.
 When Br--$B$-site bonds are strong, as in CsGeBr$_3$, octahedral tilting modes will be stiff.
In contrast, Br--$B$-site bonds are weakest in CsPbBr$_3$, and the corresponding tilting modes will be soft, while CsSnBr$_3$ will behave in a manner between the two limits but closer to CsPbBr$_3$.
The qualitative differences expected for the dynamics of octahedral tilting can be quantified by time correlation functions (TCFs) of a tilting order parameter~\cite{hyltonfarrington2024}.
We characterize fluctuations in octahedral tilting through rotation of the $B$-site--Br octahedron, which can be quantified by a rotor function that belongs to the $T_{1g}$ irrep and takes the $B$-Br vectors as input; ${\mathcal M}_{\mathcal K}^{4,m}$ with $m=3,4,5$.
The appropriate TCF is
\begin{equation}
C_{\rm tilt}(t) = \frac{ \sum_m \delta_{T_{1g} ,\tilde{\Gamma}(4,m)} \avg{\Mb_{\mathcal K}^{4,m}(0) \Mb_{\mathcal K}^{4,m}(t)}} 
{\sum_m \delta_{T_{1g} ,\tilde{\Gamma}(4,m)} \avg{\Mb_{\mathcal K}^{4,m}(0)^2}},
\label{eq:tilttcf}
\end{equation}
where $\tilde{\Gamma}(\ell,m)$ is a function that returns the irrep label, $\Gamma^{(j)}$, corresponding to each specific combination of $\ell=4$ and $m$, and $\delta_{\alpha,\beta}$ is the Kronecker delta function.
The octahedral tilting TCF, $C_{\rm tilt}(t)$, for CsGeBr$_3$ exhibits a rapid oscillatory decay with a deep minimum at small times. 
The unnormalized TCFs are also consistent with less tilting; the GeBr$_6$ octahedra do not tilt as far as the octahedra in the other two systems, as expected from the strong Br--Ge bonds, Fig.~S6.
%

 \begin{figure}[tb]
   {\includegraphics[width=0.48\textwidth]{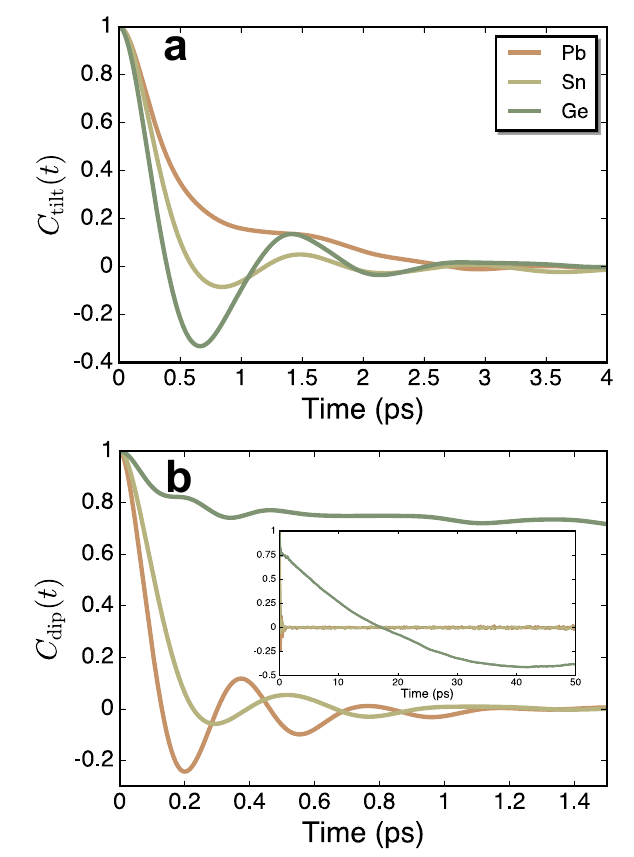}}   
  \caption{\textbf{Dynamic competition between tilting and lone pair dynamics.}
  \textbf{a} Time correlation function quantifying octahedral tilting dynamics.
  \textbf{b} Time correlation function quantifying lone pair rotational dynamics. The inset shows the slow decay of the correlation function for Ge over a time range of 50~ps.}
  \label{fig:6}
\end{figure}

%
Fluctuations in off-centering modes correspond to exchanging the identity of the (two sets of three) equivalent Br atoms around the $B$-site, breaking Br--$B$-site bonds.
The large Br--Ge bond strength makes this exchange rare and slow. 
We characterize fluctuations in off-centering through rotation of the $B$-site--MLWFC dipole, which can be quantified by a rotor function that belongs to the $T_{1u}$ irrep and takes the $B$-site--MLWFC vector as input; ${\mathcal M}_{\mathcal K}^{1,m}$ with $m=0,1,2$.
We quantify the dynamics in $B$-site dipole rotation through a TCF analogous to that for tilting,
\begin{equation}
C_{\rm dip}(t) = \frac{ \sum_m \delta_{T_{1u} ,\tilde{\Gamma}(1,m)} \avg{\Mb_{\mathcal K}^{1,m}(0) \Mb_{\mathcal K}^{1,m}(t)}} 
{\sum_m \delta_{T_{1u} ,\tilde{\Gamma}(1,m)} \avg{\Mb_{\mathcal K}^{1,m}(0)^2}}.
\label{eq:rottcf}
\end{equation}
The rotational dynamics of the Ge lone pair are extremely slow, with $C_{\rm dip}(t)$ decaying on timescales of the order of tens of picoseconds, consistent with expectations.
This extremely slow decay is in part due to the deep energy minima caused by the larger unit cell.
Using the experimental CsGeBr$_3$ unit cell
results in a faster decay of $C_{\rm dip}(t)$, on the order of 0.5-1~ps, Fig.~S5, 
though the dynamic competition between tilting and off-centering remains. 
 CsPbBr$_3$ is at the opposite extreme; it has the weakest Br--$B$-site bond of the systems studied here.
 As a result, tilting modes are soft and readily occur.
 The corresponding $C_{\rm tilt}(t)$ decays without a pronounced negative minimum. 
 Similarly, off-centering can readily occur, with the Pb ion rattling within the PbBr$_6$ octahedron. 
 This rattling is evidenced by $C_{\rm dip}(t)$, which exhibits a rapid oscillatory decay with a negative minimum at short times consistent with anticorrelation of the lone pair orientation as Pb readily moves from one off-centering configuration to another
 on this short timescale.
 Once again, CsSnBr$_3$ falls between the above two limits. 
The octahedral tilting TCF, $C_{\rm tilt}(t)$, displays a small negative minimum and small oscillations.
The lone pair rotational TCF, $C_{\rm dip}(t)$, decays more slowly than that for Pb but does exhibit some oscillations, consistent with Sn readily moving from one off-centering configuration to another. 
%

\section{Discussion}

Our combined symmetry analysis and first principles molecular dynamics simulations indicate
that $B$-site off-centering and octahedral tilting are not coupled. 
Instead, tilting and off-centering dynamically compete and are of incommensurate symmetries; 
the two modes reside on distinct crystallographic sites and transform under different irreducible representations, preventing their constructive hybridization.
Our findings challenge the common assertion that stereochemically active lone pairs drive both distortions.
However, our conclusions are supported by and extend previous work suggesting that lone pairs are not necessary for octahedral tilting~\cite{egger2024} and that $B$-site off-centering and octahedral tilting are competing distortions~\cite{radha2018distortion}. 
While stereochemical lone pair expression indeed drives $B$-site off-centering, neither lone pair expression nor off-centering couple to octahedral tilting. 
Alternatively, our results suggest an inverse correlation between lone pair activity (Pb to Sn to Ge) and octahedral tilting that is dictated by the relative stiffness of the $B$-site--halide bond, as previously hypothesized based on comparisons of relative electronegativities of the $B$-site and halide~\cite{wang2025octahedral,bechtel2018,lee2016}. 
Our results also agree with earlier work on oxide perovskites, where $B$-site ferroelectric off-centering and
antiferrodistortive octahedral rotations compete~\cite{zhong1995competing,zhong1995first,sai2000first,Wojde:2013aa,aschauer2014competition}. 
However, for very large tilts in oxide perovskites, a coupling between off-centering and tilting can emerge~\cite{aschauer2014competition,gu2018cooperative}. 
We do not exclude this possibility for some halide perovskites, though previous work found these distortions to be competitive~\cite{radha2018distortion}.
The decoupling of octahedral tilting and $B$-site off-centering modes has implications
for design and control of functional materials. 
Octahedral tilting strongly modulates orbital overlap and lattice dynamics, influencing band-edge dispersion, carrier mobility, and phonon scattering~\cite{schilcher2021significance,mayers2018,schilcher2023correlated,vonhoff2025analysis}. 
$B$-site off-centering, in contrast, governs dielectric response and polar fluctuations, which influence ferroelectricity and charge screening. 
For example, because the modes are uncoupled, one could selectively enhance or suppress each mode by varying the electronic structure of the cations in a manner that decouples and selectively tunes dielectric response (dominated by $B$-site off-centering) and dynamic disorder (octahedral tilting) to enhance carrier transport while minimizing ion conduction, ultimately enhancing performance in photophysical applications. 
We can also consider tuning cation chemistry to introduce a coupling between the $B$-site and octahedral tilting. 
If the $B$-site cation is replaced by an ion with a more complicated electron density that can couple to octahedral tilting, that coupling can be leveraged to promote or inhibit octahedral fluctuations, depending on the sign of the coupling. 
For the electron density of the $B$-site to couple to octahedral tilting, which reduces its site symmetry from $O_h$ to $C_{4v}$, the orientation of the $B$-site electron density must overlap with a $T_{1g}$ rotor function.
This requires the presence of a significant hexadecapole ($\ell=4$) in the $B$-site electron density, which is not the case for the materials studied here.
Dipolar ($\ell=1$) $B$-site cations are spectators to octahedral tilting, consistent with the propensity for halide perovskites without lone pair-containing $B$-sites to also exhibit octahedral tilting~\cite{egger2024}.
However, introduction of $f$-electron containing $B$-site cations could induce a coupling between the $B$-site electron density and octahedral tilting. 
For example, Eu$^{2+}$ or Yb$^{2+}$ are hypothesized to stiffen octahedral tilting modes due to an increased overlap in the symmetry of the $B$-site electron density and octahedral tilting distortions~\cite{tan2023lead,hyltonfarrington2024}.
This is consistent with the perspective of Papoian and Hoffmann on how orbital character and $s-p$ mixing determine extended bonding topology.
The ability of $B$-site electron density to project onto the $T_{1g}$ tilt irrep will depend on the presence of appropriate higher-order multipolar components in the local electron density (e.g. $\ell = 4$)~\cite{A.-Papoian:2000aa,papoian2001electron,ienco2001electron,PAPOIAN19988}.
Thus adding $B$-site ions whose valence states include higher angular-momentum orbitals ($f$-electron character) could in principle lead to constructive coupling between $B$-site charge anisotropy and octahedral tilting.
Reduction in octahedral tilting would reduce the effects of tilting on phonon scattering, for example, and increase the thermal conductivity of halide perovskites~\cite{Dutta:2022aa,D4CS00038B}. 
Similarly, following established concepts on electron-counting and bonding topologies~\cite{A.-Papoian:2000aa,papoian2001electron,ienco2001electron,PAPOIAN19988}, we anticipate that aliovalent doping or changes in halide identity (Cl to Br to I) that alter electron localization and $s-p$ mixing should shift the balance between off-centering and tilt stiffness in predictable ways~\cite{roy2019novel,rao2015notable,Zhou:2018aa,Chaudhary:2023aa,PRXEnergy.2.023010,Wiktor:2023aa}.
Our findings offer a symmetry-based framework for interpreting experimental observations of fluctuating polar domains, diffuse scattering patterns, and temperature-dependent structural transitions in perovskites and related materials. 
They suggest that the often-observed coexistence of polar (off-centering) and nonpolar (tilting) fluctuations in halide perovskites may arise not from cooperative modes but from competing order parameters with distinct thermodynamic origins. 
These insights extend beyond halide perovskites to the broader class of centrosymmetric lone-pair active compounds, where tuning the relative softness of bonding and tilting networks may provide a general route to control local symmetry breaking~\cite{Dutta:2022aa,acharyya2020,D4CS00038B,ubaid2024} and design materials with engineered fluctuations, enhanced phase stability, and tailored functional response.
%

\section{Simulation Details}

We performed density functional theory (DFT)-based Born-Oppenheimer ab initio molecular dynamics (AIMD) simulations of Cs$B$Br$_3$ ($B$ = Pb, Sn, Ge).
The simulations used the QUICKSTEP electronic structure module within the CP2K ab initio code~\cite{CP2K}.
QUICKSTEP utilizes a dual atom-centered Gaussian and Plane wave (GPW) basis approach to represent wavefunctions and electron density, efficiently and precisely implementing DFT.
The molecularly optimized (MOLOPT) Goedecker-Teter-Hutter (GTH) double-$\zeta$ single polarization short-ranged (DZVP-MOLOPT-SR-GTH) Gaussian basis was selected for the expansion of orbital functions.
Our simulations employed a plane-wave basis with a cutoff (Ry) and a REL\_CUTOFF (Ry) of (400, 55), (350, 50), and (400, 50) for $B$ = Pb, Sn, and Ge, respectively, to represent the electron density~\cite{vandevondele2007}.
Core electrons were represented using Goedecker-Teter-Hutter (GTH) pseudopotentials~\cite{goedecker1996}.
Exchange correlation interactions were approximated using the Perdew-Burke-Ernzerhof (PBE) generalized gradient approximation for the exchange-correlation functional~\cite{perdew1996}, as implemented in CP2K.
Grimme's D3 van der Waals correction was applied to account for the long-range dispersion interactions in the simulations~\cite{grimme2010}.

Our simulations use a $4\times4\times4$ supercell for all perovskites based on the experimentally determined unit cells, resulting in a cubic simulation cell with lengths $L=$ $23.80$~{\AA} (Pb), $23.18$~{\AA} (Sn)~\cite{lunt2018}, and $22.12$~{\AA} (Ge).
The supercell was taken to be $4\times4\times4$, because any odd multiple unit cells restrict octahedral tilting~\cite{whalley2017} and $2\times2\times2$ would result in the octahedron sharing two vertexes with the same octahedron through periodic boundaries, possibly coupling these interactions more strongly.
We equilibrated the system for at least 5~ps before performing production runs.
These equilibration simulations propagated the equations of motion with the canonical velocity rescaling (CSVR) thermostat to maintain a constant temperature of 300~K, 450~K, 550~K for B = Sn, Pb, and Ge, respectively, for the first set of simulations in the canonical (NVT) ensemble.
A second set of AIMD simulations was run so that each $B$-site has the unit cell of Pb and a temperature of 550~K.
For the calculation of dynamic properties, we performed AIMD simulations in the microcanonical (NVE) ensemble with a timestep of 0.5~fs for a total trajectory length of at least 260~ps and 60~ps  for the first and second set of simulations, respectively.
Maximally localized Wannier functions (MLWFs)~\cite{marzari2012maximally} and their centers (MLWFCs) were computed on the fly using CP2K.
The spreads of the MLWFs were minimized according to established methods~\cite{berghold2000}.

To examine collective lattice dynamics by incorporating anharmonic effects at finite temperatures, we computed phonon dispersions directly from AIMD simulations.
We used dump2phonon~\cite{kong2011} to compute the dynamical matrix directly from our AIMD simulation trajectories.
The elements of the dynamical matrix contain the force constants for the vibrational modes of the system.
In practice, this matrix is computed within dump2phonon through Green's functions, which involves Fourier transforms of atomic positions.
The eigenvalues of the dynamical matrix give the squares of the phonon frequencies for a given wave vector.
These eigenvalues were computed and the corresponding phonon dispersion curves were constructed using the phana post-processing tool~\cite{kong2011}.

\vspace{0.75cm}

\begin{center}
\textbf{ACKNOWLEDGEMENTS}
\end{center}

\begin{acknowledgements}
We acknowledge the Office of Advanced Research Computing (OARC) at Rutgers,
The State University of New Jersey
for providing access to the Amarel cluster
and associated research computing resources that have contributed to the results reported here.
\end{acknowledgements}

\vspace{1cm}

\textbf{REFERENCES}
\vspace{-0.5cm}
\bibliography{references}  

\end{document}